\documentclass[useAMS,usenatbib,usegraphicx]{mn2e}
\usepackage{aas_macros}
\usepackage{amssymb,amsmath}

\newcommand{\us}{$\mu$s}
\newcommand{\ev}[1]{E\left\{#1\right\}}

\newcommand{\mhz}{\mathrm{MHz}}
\newcommand{\khz}{\mathrm{kHz}}
\newcommand{\ms}{\mathrm{ms}}
\newcommand{\mus}{\mu\mathrm{s}}

\begin{document}

\title[Cyclic Spectral Analysis of Radio Pulsars]{Cyclic Spectral
Analysis of Radio Pulsars}
\author[P.~B.~Demorest]{P.~B.~Demorest$^{1}$\\
$^{1}$National Radio Astronomy Observatory, Charlottesville, VA 22903,
USA}

\date{Accepted 2011 June 11.  Received 2011 June 9; in original form
2011 February 17}

\pagerange{\pageref{firstpage}--\pageref{lastpage}} \pubyear{2011}

\maketitle

\label{firstpage}

\begin{abstract}

Cyclic spectral analysis is a signal processing technique
designed to deal with stochastic signals whose statistics
vary periodically with time.  Pulsar radio emission is a
textbook example of this signal class, known as {\it
cyclostationary} signals.  In this paper, we discuss the
application of cyclic spectral analysis methods to pulsar
data, and compare the results with the traditional
filterbank approaches used for almost all pulsar
observations to date.  In contrast to standard methods, the
cyclic spectrum preserves phase information of the radio
signal.  This feature allows us to determine the impulse
response of the interstellar medium and the instrinsic,
unscattered pulse profile directly from a single
observation.  We illustrate these new analysis techniques
using real data from an observation of the millisecond
pulsar B1937+21.

\end{abstract}


\section{Introduction} 
\label{sec:intro}

Pulsar radio emission produces some of the most information-rich signals
found in astronomy.  Variation exists over a wide range of time scales,
from ns-duration giant pulses \citep{hankins:crab_ns}, to tiny
decade-scale timing fluctuations that may eventually result in the first
direct detection of gravitational radiation \citep{hobbs:ipta}.
Propagation of the radio pulses through the ionized interstellar medium
(ISM) adds another layer of wavelength-dependent complexity (and
information), including dispersion, multi-path scattering, and
diffractive and refractive scintillation \citep{rickett:review}.  These
effects provide a unique tool for exploring the structure of the ionized
ISM in our galaxy on a wide range of scales.  However, they also
represent a still not fully understood source of systematic error for
high-precision pulsar timing projects \citep{hs08, crg+10}.

The complexity of the pulsar signal has led to the invention of a
variety of specialized signal processing techniques.  Currently, the
most common approach to pulsar signal analysis is to divide the
digitally-sampled baseband voltage into a number of frequency channels
using a fast Fourier transform (FFT) or autocorrelation.  The signal
power is then detected in each channel, and integrated over a short time
to produce a power spectrum estimate.  This results in well-known
tradeoffs and limitations on time and frequency resolution due to the
competing effects of dispersive smearing and inverse channel bandwidth
\citep{lorimer:book}.  A major breakthrough came with the introduction
of coherent dedispersion \citep{hankins:psp}, in which a filter is
applied within each channel pre-detection to completely remove the
effect of dispersion, for a known dispersion measure (DM).  This allows
wider channels and hence higher time resolution to be achieved, and is
commonly used for high-precision pulsar timing
\citep[e.g.,][]{hotan:msps, demorest:phd}.  This method is still bound
by the fundamental time/frequency resolution relationship $\Delta t
\Delta \nu \gtrsim 1$, with the values of $\Delta t$ and $\Delta \nu$
fixed at the time of observation.

Traditional spectral estimation procedures are based on an implicit
assumption that the input signal is stationary, or at least
approximately so over the spectrum integration timescale.  However, the
pulsar radio signal can be more accurately described as {\it
cyclostationary} -- it is a non-stationary random signal whose
statistical properties vary periodically with time \citep{rickett:amn}.
Over the past 20 years, statistical analysis techniques have been
developed specifically to deal with cyclostationary signals, the study
of which is known as cyclic spectral analysis \citep[see reviews
by][]{gardner:cyc,antoni:scd}.  These methods examine not only power
versus frequency as in conventional spectral analysis, but also the
correlation between spectral components that arises in cyclostationary
signals.  Cyclostationarity has recently been explored in a radio
astronomy context for the identification and removal of radio frequency
interference \citep{feliachi:phd}.  However, these methods have not
previously been applied to the pulsar signal itself.  Doing so provides
access to signal content that is unreachable via standard methods, and
opens the door to a variety of novel analysis techniques.

In this paper, we describe the theory and practice of applying cyclic
spectral analysis techniques to pulsar data.  Section \ref{sec:analysis}
gives a brief overview of cyclic spectral analysis and introduces the
primary quantity of interest, the cyclic spectrum and its Fourier
transforms.  In Section \ref{sec:pulsar}, we describe in detail how the
cyclic spectrum can be computed for pulsar observations, and present an
example using actual data.  Finally, Section \ref{sec:apps} shows how
the pulsar cyclic spectrum can be used to measure and correct for ISM
scattering, and discusses possible future applications of these methods.

\section{Cyclostationarity and Cyclic Spectral Analysis}
\label{sec:analysis}

We begin with a brief overview of stochastic processes and spectral
analysis, to establish terminology.  This is followed by a introduction
of the main features of cyclic spectral anlysis.  The more rigorous
treatments presented by \citet{papoulis} for stochastic processes and
\citet{gardner:book} for cyclic spectra are useful references.  While we
frame the discussion in terms of continuous-time functions, the
discretization of the following equations is straightforward.

\subsection{Standard Spectral Analysis}
\label{sec:spectra}

A general stochastic signal, $x(t)$, can largely be characterized by its
second-order statistics.  In the time domain, these are
correlations, defined here symmetrically:
\begin{equation}
C_x(t,\tau) = \ev{x(t+\frac{\tau}{2}) x^*(t-\frac{\tau}{2})}
\end{equation}
Here, the expectation value represents an average over many realizations
of the signal.  {\em Stationary} processes are those whose correlation
function depends only on the time difference $\tau$ between any two
points and not explicitly on the time $t$.  While the signals themselves
have random variation versus time, their statistics are constant.  The
signal can alternately be characterized by the power spectrum
$S_x(\nu)$, which is the Fourier transform of the correlation function
$C_x(\tau)$.  

For ergodic signals, the correlation function can be estimated from a
single realization of the signal via a time average:
\begin{equation}
\label{eqn:acf}
C_x(\tau) = T^{-1} \int_0^T x(t+\frac{\tau}{2}) x^*(t-\frac{\tau}{2}) dt
\end{equation}
This forms the basis for practical spectral analysis of random signals.
In practice, both ergodicity and stationarity are almost always assumed
a priori, since one usually has only one realization of a signal.
Due to the Fourier-pair uncertainty relations, the spectral resolution
$\Delta\nu$ is limited to approximately $T^{-1}$ or larger.  

In radio astronomy, $x(t)$ is the baseband voltage, a quantity
proportional to the incident radio wave's electric field, mixed
(frequency-shifted) to near zero frequency.  Early digital radio
spectrometers often explicitly evaluated Eqn.~\ref{eqn:acf} using
dedicated hardware.  In modern systems, $S_x(\nu)$ is usually computed
using a fast Fourier transform (FFT) of $x(t)$.  In order for a pulse to
be resolved in time, $T$ must be small compared to the pulse width.  For
pulsars, this puts a limit on the finest achievable frequency resolution
of $\Delta \nu \gtrsim (P\Delta\phi)^{-1}$ where $P$ is the pulse period
and $\Delta \phi$ is the pulse width in turns.

\subsection{Cyclic Spectral Analysis}
\label{sec:cyclo}

An interesting class of nonstationary signal are those whose statistics
vary periodically with time.  These are known as {\em cyclostationary}
signals and their correlations exhibit periodicity:
\begin{equation}
\label{eqn:periodic_corr}
C_x(t, \tau) = C_x(t + P, \tau) = C_x(t \bmod P, \tau)
\end{equation}
It is important to note that although $C_x(t,\tau)$ is periodic in $t$,
it is not generally periodic in $\tau$, and $x(t)$ itself is not
necessarily a periodic function.  Simple examples of cyclostationary
signals can be obtained by periodically amplitude- or
frequency-modulating a stationary noise process.  In this paper we
will sometimes use the quantity pulse phase $\phi = t/P$ in place of $t$
as an independent variable in expressions such as $C_x(\phi,\tau)$.

The cyclic analog of the power spectrum can be obtained by Fourier
transforming $C_x(t,\tau)$ along both the $t$ and $\tau$ axes:
\begin{equation}
S_x(\nu;\alpha) = P^{-1} \int\limits_{-\infty}^{+\infty} d\tau 
  \int\limits_0^P dt \; 
  C_x(t,\tau) 
  e^{-2\pi i \nu \tau} e^{-2\pi i \alpha t}
\end{equation}
This quantity is known as the {\em cyclic spectrum} and it is a function
of two frequency variables.  The frequency $\nu$ is conjugate to $\tau$
exactly as in the standard power spectrum.  The {\em cycle frequency}
$\alpha$ is conjugate to $t$, and takes on the discrete values $\alpha_n
= n/P$ -- it is computed here as a Fourier series rather than a
continuous Fourier transform.  In the stationary signal case,
$S_x(\nu;\alpha)=0$ for $\alpha\neq0$.

The periodic correlation can be estimated from a single signal
realization similarly to Eqn.~\ref{eqn:acf} by ``folding'' the
correlations modulo the periodicity $P$:
\begin{equation}
\label{eqn:corrsum}
C_x(t,\tau) = N^{-1} \sum_{n=0}^N x(t+nP+\frac{\tau}{2})
  x^*(t+nP-\frac{\tau}{2})
\end{equation}
This again assumes ergodicity of the signal, over $N$ periods.  In
this computation, the integration time $T$ is given by $NP$, the number
of periods folded.  The integration time sets an ultimate limit on
the range of $\tau$ that can be measured, and again the spectral
resolution is limited to $\Delta \nu \gtrsim T^{-1}$.  However, in
constrast with Eqn.~\ref{eqn:acf}, $T$ is not constrained by the pulse
width, as the nonstationarity has already been correctly taken into
account in Eqn.~\ref{eqn:corrsum}.  This decouples the pulse phase
resolution from the frequency resolution.

It is easy to see that various symmetry relationships exist among these
quantities.  In our symmetric formulation, $C_x(t,-\tau)=C^*_x(t,\tau)$.
Similarly, $S_x(\nu;-\alpha)=S^*_x(\nu;\alpha)$.  A useful intermediate
quantity is the {\em periodic spectrum} $S_x(\nu,t)$ that is obtained by
Fourier transforming $C_x(t,\tau)$ with respect to $\tau$ alone.  By
symmetry, $S_x(\nu,t)$ is purely real-valued, and therefore is
straightforward to visualize.  $S_x(\nu,t)$ is superficially similar to
signal power as a function of time and frequency.  However as discussed
below, it also contains information about the signal phase, and thus is
not strictly positive.

It can be shown that the cyclic spectrum also represents correlations in
the frequency domain \citep{gardner:cyc, gardner:book}.  Using the
Fourier transform of $x(t)$, $X(\nu)$, we have:
\begin{equation}
\label{eqn:fcorr}
S_x(\nu;\alpha) = \ev{X(\nu + \alpha/2) X^*(\nu - \alpha/2)}
\end{equation}
This relationship directly leads to a key result on which we will rely
heavily in later sections:  If $y(t)$ is the result of passing $x(t)$
through a linear, time-invariant filter with impulse response $h(t)$
(frequency response $H(\nu)$),
\begin{equation}
y(t) = h(t) \star x(t) 
\end{equation}
\begin{equation}
Y(\nu) = H(\nu) X(\nu)
\end{equation}
then the cyclic spectra of $x$ and $y$ are related by:
\begin{equation}
\label{eqn:inout}
S_y(\nu;\alpha) = H(\nu+\alpha/2) H^*(\nu-\alpha/2) S_x(\nu;\alpha)
\end{equation}

Eqn.~\ref{eqn:inout} shows that the phase content of $H(\nu)$ is
preserved in the cyclic spectrum, whereas in a standard power spectrum
the only information retained is the filter magnitude
$\left|H(\nu)\right|^2$.  This is a fundamental difference between the
two approaches that will enable most of the novel applications described
in \S\ref{sec:apps}.

In practice, computation of any of the cyclic spectrum variants will be
done using band-limited, digitized sampled values.  That is, the signal
from the radio frequency of interest will be filtered with a bandpass
filter of width $B$, mixed to baseband and critically sampled at the
Nyquist rate.  It is well known that any remaining power outside the
baseband frequency range $-B/2<\nu<B/2$ will be aliased into the sampled
values.  Similar considerations applied to cyclic spectra show that the
resulting $S(\nu;\alpha)$ is only valid within a diamond-shaped region
around the origin, defined by $|\alpha/2| + |\nu| < B/2$.

\section{Implementation for Pulsar Data}
\label{sec:pulsar}

A simple but useful model of the pulsar radio signal $y(t)$ as it is
received on Earth is amplitude-modulated noise
\citep{rickett:amn}\footnote{For simplicity, the model presented here
neglects the complex ``subpulse'' behavior of many radio pulsars as well
as other filtering processes caused by the ionosphere, troposphere,
telescope equipment, etc.  This does not alter our basic conclusions.}: 
\begin{equation}
  y(t; t_0, \nu_0) = h_{ISM}(t; t_0) \star \left( p(t\Omega(t_0); \nu_0)
  n_i(t) \right)
\end{equation}
Here, $n_i(t)$ is a stationary noise process representing the intrinsic
pulsar radio emission.  $p(\phi; \nu_0)$ is a periodic function
describing the average pulse profile shape at a radio frequency $\nu_0$.
The timing model $\Omega(t_0)$ gives the apparent rotational 
frequency of the pulsar at a given time.  This includes the
slowly-changing intrinsic spin frequency of the pulsar as well as
Doppler and relativistic terms due to the motions of the Earth and
pulsar.  The impulse response function, $h_{ISM}(t; t_0)$, describes the
effect of the ionized interstellar medium (ISM) on the radio signal.  It
contains both dispersion and scintillation/scattering terms, and slowly
evolves with a characteristic timescale $t_d$ due to the relative
motions of the Earth, ISM, and pulsar.  For clarity, we will often omit
the slow dependences on $t_0$ and $\nu_0$ in the following discussion.

Under the approximation that the average intrinsic pulsar emission
spectrum $S_n(\nu)=S_0$ is flat (valid over small fractional
bandwidths), the pre-ISM pulsar cyclic spectrum takes the simple form
$S_x(\nu;\alpha_n) = I(n)S_0$, where $I(n)$ is the Fourier transform of
the intensity profile $I(\phi)=\left[p(\phi)\right]^2$ and $\alpha_n =
n\Omega$.  The addition of the ISM filtering process produces an
expression for the final pulsar cyclic spectrum:
\begin{equation}
\label{eqn:model}
  S_y(\nu; \alpha_n) = 
    H_{ISM}(\nu+\frac{\alpha_n}{2}) H^*_{ISM}(\nu-\frac{\alpha_n}{2})
    I(n) S_0
\end{equation}
The duration of the ISM impulse response, $\tau_{ISM}$, can be
approximated as the combination of the two primary effects,
dispersion and scattering: $\tau_{ISM}^2 \approx \tau_{DM}^2 +
\tau_{sc}^2$.  In order to resolve the full frequency structure of
$H_{ISM}(\nu)$, the cyclic spectra must be computed with a frequency
resolution $\Delta \nu \lesssim (2\tau_{ISM})^{-1}$.  Since
$H_{ISM}(\nu)$ is slowly evolving with time, the cyclic spectrum
integration time should be kept shorter than the diffractive
scintillation timescale $t_d$ for the signal model presented in
Eqn.~\ref{eqn:model} to apply.  

\subsection{Time-domain Approach}
\label{sec:time}

A time domain approach to computing the cyclic spectrum is based on
directly evaluating Eqn.~\ref{eqn:corrsum}.  In this method, the
complex-conjugated baseband values are delayed by an integer number of
samples (``lags''), and then multiplied by the original undelayed
version.  The pulse phase for each point in this cross-multiplied data
stream is calculated using standard methods (polynomial expansion of
$\Omega(t_0)$) and then they are folded into the desired number of pulse
phase bins.  The process is repeated for the desired number of delays
($N_{lag}$).  Due to symmetry, only positive lags need be evaluated.
When combined with the ``mirrored'' negative lag values, the result is
an $N_{bin}$ by $N_{chan} = 2N_{lag}-1$ array representation of
$C_x(\phi,\tau)$.  This computation can be preceded by standard coherent
dedispersion at a known DM to reduce the required frequency resolution,
effectively setting $\tau_{DM}$ to $\sim0$.

This time domain method has the benefit of straightforward handling of
the relationship between the slowly changing pulse period and sampling
rate.  It also is easily integrated into standard pulsar observing
practices:  The ``zero-lag'' term is simply a standard folded pulse
profile as would usually be computed by a traditional pulsar processing
system.  The drawback to this method is that the computational cost per
sample scales linearly with $N_{chan}$, quickly becoming impractical if
fine frequency resolution is desired.  For real time computation
over a total bandwidth $B$ with frequency resolution $\Delta \nu$, the
computational cost in floating-point operations per second (flops) is
given by
\begin{equation}
\label{eqn:td_cost}
  \mathcal{C}_{T}(B,\Delta\nu) \simeq 4 N_{pol} \frac{B^2}{\Delta \nu}
\end{equation}
Here, $N_{pol}$ equals 2 if only total power terms are computed, or
4 if polarization cross-products are included.  An implementation of the
time-domain cyclic spectrum method is now included in the open-source
{\sc DSPSR} software package\footnote{http://dspsr.sourceforge.net}
\citep{dspsr}.

\subsection{Frequency-domain Approach}
\label{sec:freq}

\begin{figure}
\begin{center}
\includegraphics[width=0.47\textwidth]{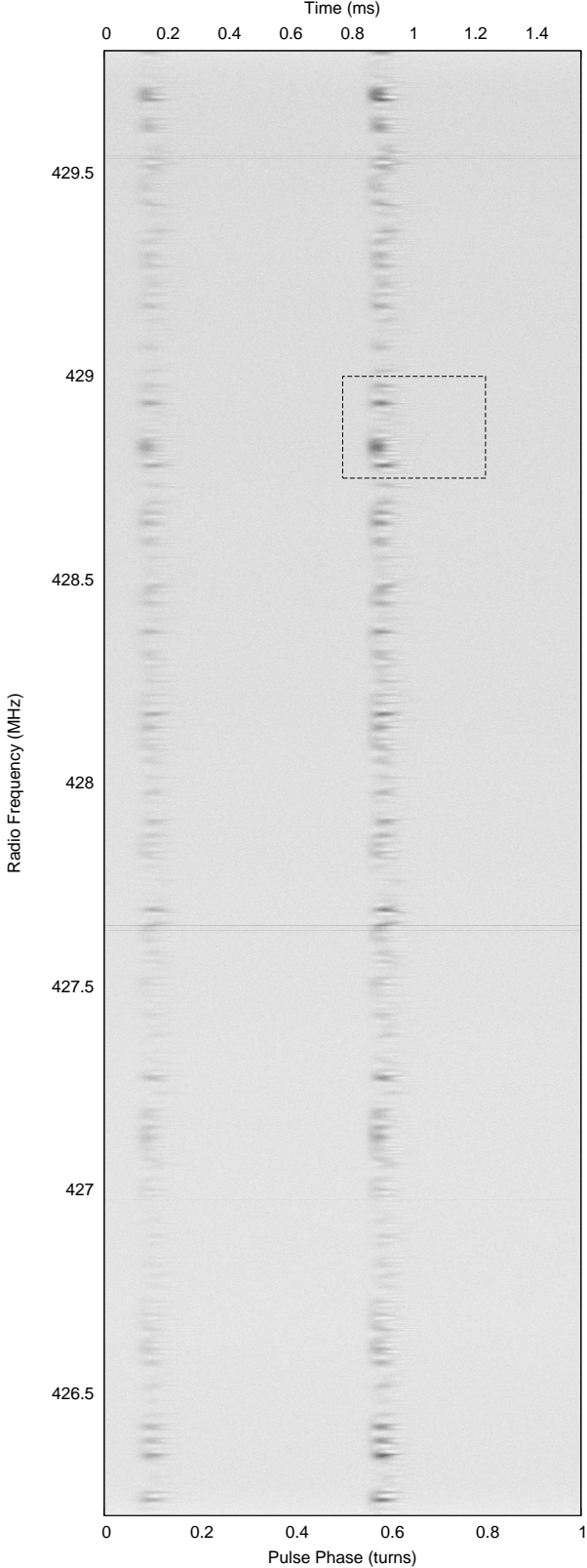}
\caption{\label{fig:ps} Periodic spectrum $S_x(\nu,t)$ of B1937+21 using
frequency resolution of $4~\mhz / 6320 = 0.642~\khz$ and pulse phase
resolution of $1.55~\ms / 511 = 3~\mus$.  The boxed region is shown in
detail in Figure~\ref{fig:ps_zoom}.}
\end{center}
\end{figure}

It is also possible to compute the cyclic spectrum in the frequency
domain, based on Eqn.~\ref{eqn:fcorr}.  In this method, the raw samples
are first Fourier transformed with resolution determined by the
$\tau_{ISM}$ timescale.  The frequency domain data is multiplied by an
appropriate phase gradient to align the data in pulse phase, and a
coherent dedispersion filter is optionally applied at this point.  The
frequency domain correlations of Eqn.~\ref{eqn:fcorr} are then computed
for $N_{lag}^\prime$ ``spectral lags'' of spacing $\alpha_n$.  The
number of spectral lags determines the pulse phase resolution, with
$N_{bin}=2N_{lag}^\prime - 1$.  However, since there is generally not an
integer number of pulse periods per FFT length, the frequencies required
for evaluating Eqn.~\ref{eqn:fcorr} fall between FFT bins.  In order to
avoid artifacts in the resulting cyclic spectra, the FFT output must be
interpolated to obtain the correct fractional frequencies.  Here we use
a 7-point sinc interpolation with good results and leave a detailed
study of the effect of this interpolation for future work.  Finally the
correlations are averaged into $N_{chan}$ frequency bins.  As in the
time domain case, applying coherent dedispersion reduces the final
required frequency resolution.

The frequency domain method has the advantage of the computational
requirements scaling linearly with $N_{bin}$ rather than $N_{chan}$ as
in the time domain version.  This is attractive for situations where
very fine frequency resolution is desired.  However, there is some added
complexity due to the frequency interpolation, and this method is a much
larger departure from standard pulsar observing strategies.  For
this method, the number of flops required for real-time computation is
\begin{equation}
\label{eqn:fd_cost}
  \mathcal{C}_{F}(B,\Delta\nu) \simeq N_{bin} (2N_{int} + 4N_{pol})
  B + 5 B \log_2 \frac{B}{\Delta \nu}
\end{equation}
where $N_{int}$ is the number of points in the interpolation kernel.
The second term is the approximate cost of the initial FFT.  Real-world
FFT performance can depart significantly from this scaling depending on the
implementation details and computing hardware architecture.

\begin{figure*}
\begin{center}
\includegraphics[width=\textwidth]{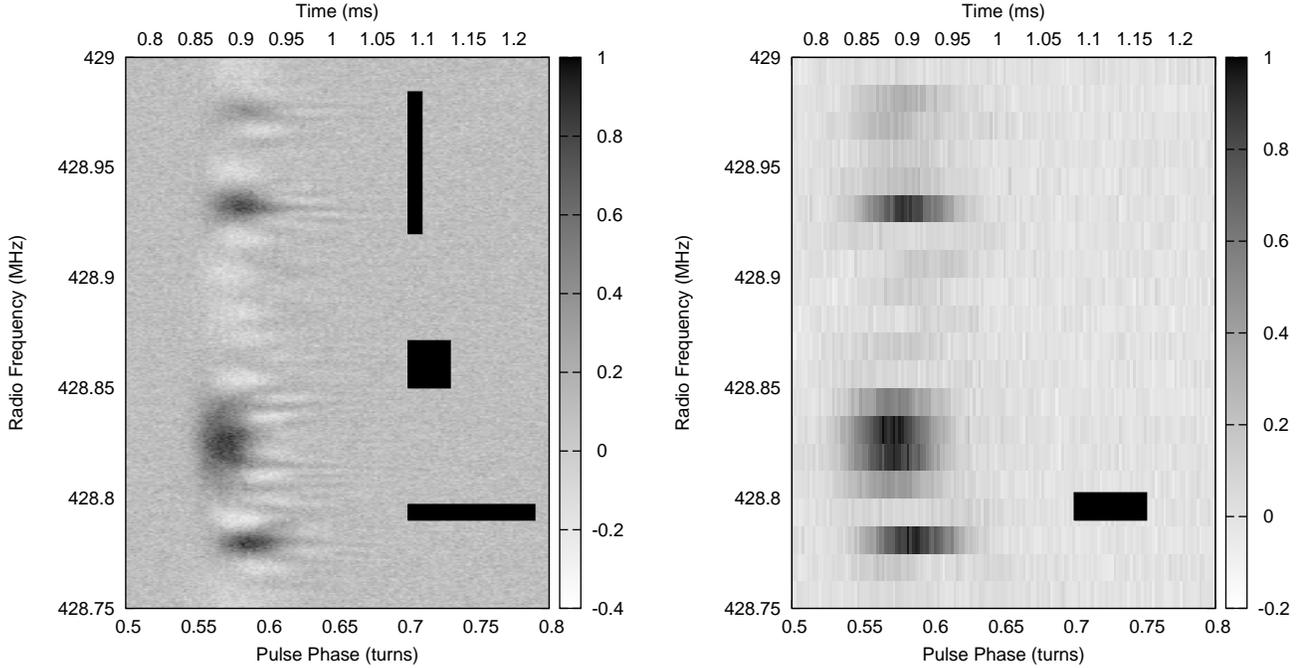}
\caption{\label{fig:ps_zoom} Left: Detail of periodic spectrum of
B1937+21.  The black rectangles each contain unity time-frequency area.
The fixed resolution implied by any one of these, as is produced by a
traditional filterbank, would miss various features in the data.  Note
also the appearance of both positive (dark) and negative (light) signal
regions at fine scales.  This is due to the phase content of
$H_{ISM}(\nu)$ present in this representation.  Right: The same data
processed with a standard coherently dedispersed filterbank \citep[{\sc
DSPSR};][]{dspsr}, using a frequency resolution of 12.5~kHz, and
corresponding intrinsic time resolution of 80~\us\ (black rectangle).
The data are oversampled in pulse phase by a factor of $\sim$20,
resulting in the same number of profile bins as in the left panel.  In
both panels the zero-point has been determined by measuring the mean
off-pulse level, and the data have been normalized by the peak value.}
\end{center}
\end{figure*}

\subsection{Hybrid Filterbank Methods}
\label{sec:filterbank}

Neither the time-domain method of Section~\ref{sec:time} or the
frequency-domain method of Section~\ref{sec:freq} is efficient for
direct application to a wide ($B \gtrsim 10$~MHz) frequency band.
$\mathcal{C}_{T}$ scales as $B^2$, and while $\mathcal{C}_{F}$ formally
grows only as $B\log B$, the FFT length required for dedispersion grows
as $B^2$ and eventually becomes impractical in current computing
hardware.  The computational burden can be reduced by first splitting
the wide band into a number of sub-bands via a filterbank, and then
applying either the time or frequency domain cyclic spectrum computation
method in each sub-band separately.  This is similar to how most
coherent dedispersion algorithms currently work.  However, in contrast
with standard filterbanks, to compute a continuous cyclic spectrum over
the full wide band, the sub-bands must be made to {\it overlap in
frequency}.  As explained at the end of Section \ref{sec:cyclo}, the
cyclic spectrum can not be properly computed for frequencies within
$\alpha/2$ of an edge of a band-limited signal.  Therefore to avoid gaps
at each sub-band edge, the sub-bands must overlap by an amount $B_{ol}
\geq \alpha_{max}$, and the invalid edges discarded before finally
appending the sub-band results together.  This approach results in a
total number of real-time flops given by
\begin{equation}
\label{eqn:hybrid_cost}
  \mathcal{C}_H \simeq \mathcal{C}_{FB}(B,B_s) 
    + \frac{B}{B_s - B_{ol}} \mathcal{C}_{T,F}(B_s,\Delta \nu)
\end{equation}
where $\mathcal{C}_{FB}$ is the cost of the filterbank operation,
$B_s$ is the full width of each sub-band (including the overlapped
portion), and $\mathcal{C}_{T,F}$ is the cost of either of the
previously explained methods applied to a single sub-band.  Due to the
relationship $B_s > B_{ol} \geq \alpha_{max}$, the minimum possible
sub-band width is set by the desired pulse phase resolution $\Delta \phi
= (2P\alpha_{max})^{-1}$. 

\subsection{Example Data}
\label{sec:example}

We illustrate the concepts just discussed with an example of cyclic
spectrum computation on actual pulsar data.  We observed the 1.55-ms
pulsar B1937+21 at Arecibo Observatory\footnote{The Arecibo Observatory
is part of the National Astronomy and Ionosphere Center, which is
operated by Cornell University under a cooperative agreement with the
National Science Foundation.} using the Astronomy Signal Processor (ASP)
pulsar backend \citep{demorest:phd}.  Dual-polarization, 8-bit baseband
voltage samples were recorded in a 4~MHz band centered on 428~MHz.
Cyclic spectra were computed via the frequency domain
method\footnote{The same data analyzed with the time domain method
produced identical results.} discussed in \S\ref{sec:freq} using a FFT
length of 542288 points, 256 spectral lags and integrated for
$T_{int}=2$~minutes.  The FFT length was chosen based on the dispersion
time $\tau_{DM}\sim30$~ms.  A coherent dedispersion filter using a DM of
71.019~pc~cm$^{-3}$ was applied before summing the cyclic spectra to the
final output frequency resolution of $N_{chan}=6230$
($\Delta\nu=0.642$~kHz).  When transformed to the periodic spectrum
domain (Figures~\ref{fig:ps} and \ref{fig:ps_zoom}), the 256 harmonics
result in $N_{bin}=511$ ($P\Delta\phi=3.0$~\us).  The spectra were
computed separately for each polarization, then summed to make the total
intensity spectrum shown.  For this example, the radio frequency
resolution $\Delta \nu$ was chosen to be approximately equal to $1/P$,
providing equal spacing between points in the $\nu$ and $\alpha$
dimensions in the cyclic spectrum.

PSR~B1937+21 is an ideal case for showing the utility of these methods.
Near 430~MHz, the typical ISM scattering time is $\tau_{sc}\sim40~\mus$,
comparable to the intrinsic pulse width \citep{cordes:1937}.  Even if
coherent dedispersion is first performed, the typical filterbank
approach forces one to choose between: 1. Resolving the scintillation
structure (on frequency scales $\sim\tau_{sc}^{-1}$) but losing almost
all pulse shape information; or 2. Retaining enough time resolution to
resolve the pulse, but missing most of the fine frequency structure.
These tradeoffs are graphically illustrated in Figure~\ref{fig:ps_zoom},
where it can be seen that the cyclic spectrum approach retains
information on both small pulse phase and frequency scales.  As
previously mentioned, the periodic spectrum contains signal phase
information and the on-pulse region can be seen to contain both positive
and negative components, relative to the mean off-pulse level.  Also
notable are the fine ``streaks'' that appear at later pulse phases.  One
possible explanation for these features is that in this part of the
profile the signal is dominated by more highly delayed components,
making the periodic correlation $C(\phi,\tau)$ contain signal power
at larger $|\tau|$ values than are present at earlier phases.  When
transformed to the periodic spectrum domain, this results in finer frequency
structure appearing at later pulse phases.

\section{Applications}
\label{sec:apps}

\begin{figure}
\begin{center}
\includegraphics[width=0.47\textwidth]{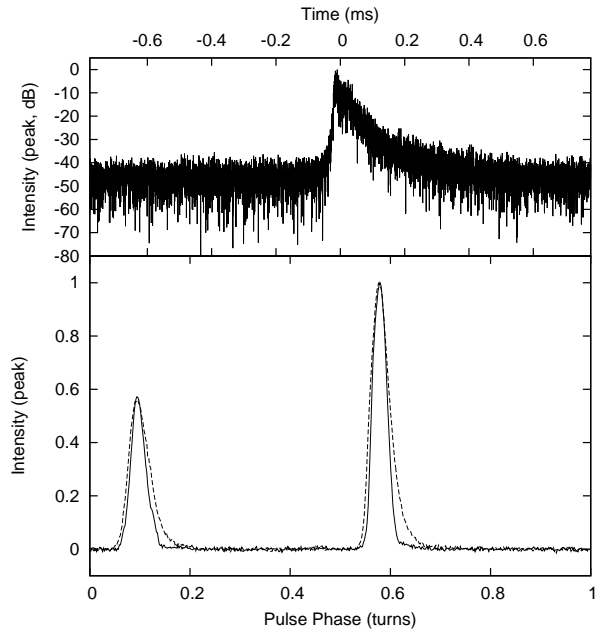}
\caption{\label{fig:profs} Top: Absolute square of the impulse response
$|h_{ISM}(t)|^2$ determined from the data shown in Figure~\ref{fig:ps}.
Bottom: Corresponding standard (dashed) and scattering-corrected (solid)
pulse profiles.  All functions have been normalized by their peak value.
The x-axis range covers the same total time span in both panels.}
\end{center}
\end{figure}

Given the signal model presented in Eqn.~\ref{eqn:model}, it is possible
to determine both the ISM response and intrinsic pulse profile directly
from a single cyclic spectrum.  The two-dimensional cyclic spectrum
$S(\nu;\alpha_n)$ contains $N_{chan} \times N_{lag}^\prime$ data values,
while $H_{ISM}(\nu)$ and $I(n)$ are described by only
$N_{chan}+N_{lag}^\prime$ model parameters. This provides sufficient
constraints for both to be determined via iterative least-squares
minimization.  A detailed analysis of this method will be presented
in a separate paper \citep{walker:cyc}.  One previous measurement
method for $h_{ISM}(t)$ has been published \citep{walker:hol}, based on
a dynamic spectrum phase retrieval procedure.  The cyclic spectrum
method is much simpler since in this case the wave phases can be
measured directly, giving an estimate of $h_{ISM}$ from a single
``snapshot'' observation.  The cyclic spectrum also incorporates pulse
profile shape information, which is lost in standard dynamic spectra.
This naturally leads, for the first time, to a true coherently
descattered pulse profile shape (Figure~\ref{fig:profs}).  In contrast
with previous intensity-based profile deconvolution methods
\citep{bhat:pbf}, this requires no assumption of a specific functional
form for $h_{ISM}$ and is not affected by ambiguity between intrinsic
and ISM-induced profile features.  The ISM response shown in
Figure~\ref{fig:profs} has an initial exponential decay followed by a
more slowly-decaying tail.  This will be interesting to compare in
detail with the predictions of the standard Kolmogorov scattering model
\citep[e.g.,][]{rjt+09}.

There are several degeneracies that the descattering process alone can
not resolve.  As is clear from Eqn.~\ref{eqn:model}, multiplying
$H_{ISM}$ by a constant phase factor will produce no change in the
observed cyclic spectrum.  Similarly, without additional assumptions,
the pulsar's intrinsic flux $S_0$ is degenerate with the magnitude of
$H_{ISM}$.  Most critically for pulsar timing, an arbitrary rotation can
be applied to $I(\phi)$, and absorbed into $H_{ISM}$.  That is, at this
level of analysis it is impossible to distinguish an ISM-induced delay
from a pulsar spin deviation or other timing effect.  Resolving this
situation to obtain properly scattering-corrected timing will require
the development of additional analysis techniques.  This could
range from assuming a constrained form or applying moment analysis to
$h_{ISM}$ to physical models of the spatial distribution of scattering
material, and is an active topic for further study.  Compared with
previous methods, the cyclic spectrum provides a qualitatively new way
to measure the ISM response, and this new information dramatically
expands the possibilities for scattering corrections to timing data.

In addition to the determination of $H_{ISM}$, Eqn.~\ref{eqn:inout}
allows for the applcation of other phase-coherent filters to the final
integrated cyclic spectra.  This technique can be used to perform
coherent dispersion corrections, within the limits of the cyclic
spectrum resolution.  The high frequency resolution of the pulsar cyclic
spectrum enables precise post-detection removal of narrow-band radio
frequency interference, without sacrificing pulse phase resolution.
More sophisticated approaches may also incorporate information gained
from the cyclostationarity of the interference \citep{feliachi:phd}.

Although the discussion in \S\ref{sec:analysis} focused on the analysis
of a single stochastic signal, pairs of correlated cyclostationary
signals can be analyzed via cyclic cross-spectra, in complete analogy
with standard cross-spectra \citep{gardner:book}.  For dual-polarization
radio data, this results in the creation of cyclic Stokes parameters,
and allows for the application of phase-coherent matrix convolution
\citep{straten:phase} directly to the cyclic spectra.  For radio
interferometers, analyzing data from antenna pairs will produce cyclic
visibilities.  Along with recently developed VLBI imaging techniques for
investigating pulsar scintillation \citep{brisken:vlbi_arcs}, this could
prove to be an extremely powerful tool for exploring the ISM.

\section{Conclusions}
\label{sec:conclusion}

Cyclic spectral analysis is a powerful new observational technique for
studying radio pulsars.  It provides a data representation that
simultaneously preserves both high pulse phase resolution and high
frequency resolution information about the signal.  With the
accompanying preservation of signal phase content, this allows
fundamentally new analysis techniques not possible with standard
filterbank data.  This holds promise both for increasing our
understanding of the ionized ISM, and eventually for removing ISM
scattering as an obstacle to achieving the highest possible pulsar
timing precision.

\section*{Acknowledgements}
During the course of this research, P.B.D. has received support from
both a Jansky Fellowship of the National Radio Astronomy Observatory,
and the NSF PIRE program award number 0968296.  I would like to thank
Mark Walker and Willem van Straten for much discussion and support
during the development of these ideas, and the referee, Barney Rickett,
for his helpful comments.

\bibliographystyle{mn2e}
\bibliography{cyclic}

\label{lastpage}

\end{document}